\documentclass[a4paper,fleqn,usenatbib]{mnras}

\usepackage{newtxtext,newtxmath}

\usepackage[T1]{fontenc}
\usepackage{ae,aecompl}

\usepackage{graphicx}	
\usepackage{amsmath}	
\usepackage{amssymb}	
\usepackage{ragged2e}

\newcommand{\hii}{\ion{H}{ii} }
\newcommand{\nii}{\ion{N}{ii} }
\newcommand{\oiii}{\ion{O}{iii} }
\newcommand{\oii}{\ion{O}{ii} }
\newcommand{\sii}{\ion{S}{ii} }

\title[Arm-interarm gas abundance variations with MUSE]{Arm-interarm gas abundance variations explored with MUSE: the role of spiral structure in the chemical enrichment of galaxies}

\author[L. S\'anchez-Menguiano et al.]{
Laura S\'anchez-Menguiano,$^{1,2}$\thanks{E-mail: lsanchez@iac.es}
Sebasti\'an F. S\'anchez,$^{3}$
Isabel P\'erez,$^{4,5}$ Tom\'as 
\newauthor Ruiz-Lara,$^{1,2}$ 
Llu\'is Galbany,$^{4}$
Joseph P. Anderson,$^{6}$
and Hanindyo Kuncarayakti$^{7,8}$
\\
$^{1}$Instituto de Astrof\'isica de Canarias, La Laguna, Tenerife, E-38200, Spain\\
$^{2}$Departamento de Astrof\'isica, Universidad de La Laguna, Tenerife, Spain\\
$^{3}$Instituto de Astronom\'ia, Universidad Nacional Aut\'onoma de M\'exico, A.P. 70-264, C.P. 04510, M\'exico D.F., Mexico\\
$^{4}$Dpto. de F\'isica Te\'orica y del Cosmos, Universidad de Granada, Facultad de Ciencias (Edificio Mecenas), E-18071 Granada, Spain\\
$^{5}$Instituto Carlos I de F\'isica Te\'orica y computacional, Universidad de Granada, E-18071 Granada, Spain\\
$^{6}$European Southern Observatory, Alonso de C\'ordova 3107, Casilla 19 Santiago, Chile\\
$^{7}$Tuorla Observatory, Department of Physics and Astronomy, FI-20014 University of Turku, Finland\\
$^{8}$ Finnish Centre for Astronomy with ESO (FINCA), FI-20014 University of Turku, Finland
}

\date{Accepted 2019 January 9. Received 2019 December 20; in original form 2019 October 17}

\pubyear{2020}

\begin{document}
\label{firstpage}
\pagerange{\pageref{firstpage}--\pageref{lastpage}}
\maketitle

\begin{abstract} 
Spiral arms are the most characteristic features of disc galaxies, easily distinguishable due to their association with ongoing star formation. However, the role of spiral structure in the chemical evolution of galaxies is unclear. Here we explore gas-phase abundance variations between arm and interarm regions for a sample of 45 spiral galaxies using high spatial resolution VLT/MUSE Integral Field Spectroscopy data. We report the presence of more metal-rich \hii regions in the spiral arms with respect to the corresponding interarm regions for a large subsample of galaxies ($45-65\%$ depending on the adopted calibrator for the abundance derivation). A small percentage of the sample is observed to display the opposite trend, that is, more metal-poor \hii regions in the spiral arms compared to that of the interarms ($5-20\%$ depending on the calibrator). We investigate the dependence of the variations with three galaxy properties: the stellar mass, the presence of bars, and the flocculent/grand design appearance of spiral arms. In all cases, we observe that the arm-interarm abundance differences are larger (positive) in more massive and grand-design galaxies. This is confirmed by an analogous spaxel-wise analysis, which also shows a noticeable effect of the presence of galactic bars, with barred systems presenting larger (positive) arm-interarm abundance variations than unbarred systems. The comparison of our results with new predictions from theoretical models exploring the nature of the spirals would highly impact on our knowledge on how these structures form and affect their host galaxies. 
\end{abstract}




\section{Introduction}\label{sec:intro}
The study of the gas-phase chemical composition of spiral galaxies has proven to be a powerful tool to improve our knowledge on the evolution of these complex systems. In particular, the analysis of \hii regions is of great importance, as it is through the birth and death of stars that galaxies chemically evolve. 

The most studied trend in the oxygen abundance distribution of \hii regions in spiral galaxies is the well-known negative radial gradient \citep[e.g.][among many others]{searle1971, martin1992, sanchez2012b, ho2015, belfiore2017, sanchezmenguiano2018, zinchenko2019}. Observational works on this topic have enabled constraints on chemical evolution models aimed at explaining galaxy formation and evolution \citep{edmunds1995, molla1997, prantzos2000, chiappini2001, molla2005, pilkington2012}. Nevertheless, by restricting the information to one dimension they were unable to properly describe the effect of two-dimensional (2D) structures, such as spiral arms or bars, on the chemical evolution of their host galaxies. 

From a theoretical point of view, numerical simulations have recently revealed how the presence of spiral arms produces azimuthal differences in the chemical distribution \citep{dimatteo2013, grand2016, sanchezmenguiano2016b}. One proposed mechanism responsible for these differences is the redistribution of metals induced by streaming motions of gas and stars along the spiral arms. However, it has also been concluded that the passage of spiral density waves across galactic discs alone can cause azimuthal metallicity variations \citep{ho2017, molla2019, spitoni2019}.

Observationally, one of the most straightforward methods to study the effect of the spiral structure on the chemical distribution of galaxies is via the analysis of arm-interarm abundance variations. However, the first attempts to carry out such study found no significant differences in the gas metallicity between these two distinct regions \citep{martin1996, cedres2002}. More recently, \citet{sanchezmenguiano2017} analysed a sample of 63 galaxies from the CALIFA survey \citep{sanchez2012a} and reported very subtle differences between the arm and interarm abundance gradients for two specific subgroups of galaxies: barred and flocculent\footnote{Flocculent galaxies are defined as those displaying short, asymmetric, and patchy arms that fade over the gaseous disc \citep{elmegreen1981}. See Sec.~\ref{sec:properties} for more information on this morphological classification of spiral arms.} systems. \citet{sakhibov2018} also analysed four CALIFA spiral galaxies, showing a very small enhancement ($0.01-0.06$ dex) of the oxygen abundance in the spiral arms compared to the interarm region.

A drawback of those previous studies - that could account for the absence of strong evidence of arm-interarm abundance variations - is the lack of the spatial resolution of the data that might be needed to detect such differences. In this regard, the advent of Integral Field Spectroscopy (IFS) instruments combining large field-of-views (FoV) and high spatial resolution has meant a revolution in the progress of 2D abundance distribution studies. Since then, a growing number of works have been able to detect such elusive variations in individual galaxies: NGC~6754 (\citealt{sanchezmenguiano2016b}, see also \citealt{sanchez2015a}), NGC~1365 \citep{ho2017}, HCG~91c \citep{vogt2017}, and NGC~2997 \citep{ho2018}. The magnitude of the reported variations differ from study to study (from 0.06 in NGC~2997 to 0.4 dex in NGC~1365), but in all cases more metal-rich gas has been observed associated with the spiral structure. 

In spite of the improvement in the spatial resolution of the available data, there are still studies reporting an absence of arm-interarm oxygen abundance variations. \citet{kreckel2016} analysed MUSE \citep[Multi Unit Spectroscopic Explorer;][]{bacon2010} data of NGC~628 and detected 391 \hii regions covering both arm and interarm regions, deriving a metallicity distribution independent of the environment. However, the data did not map the total extent of the galaxy, but were restricted to a region of 12 kpc$^2$, barely 10\% of the total area\footnote{Assuming $R_{25}=5.25$ arcmin \citep{kendall2011}.}. It is therefore not clear that strong conclusions can be drawn when such a small percentage of the galaxy was analysed.

A first attempt to study azimuthal oxygen abundance variations in an homogeneous way for a number of galaxies has been recently performed by \citet{kreckel2019}. Analysing a sample of eight spiral galaxies from the PHANGS-MUSE project, the authors qualitatively find subtle azimuthal variations tenuously associated with the spiral pattern in half of the sample. This spatial overlapping with the spiral structure was more clearly observed in some galaxies (such as NGC~1087 or NGC~1672, where enhanced abundances were detected along the eastern arm) than in others. In general, they find stronger correlations with local physical conditions of the ISM (especially with the ionisation parameter) than with environmental parameters (arm/interarm masks or stellar mass surface density offset). The authors conclude that the spiral arms can play an important role in mixing the ISM, in combination with a complex interplay between galaxy dynamics and enrichment patterns.

In order to shed light on the prevalence of the arm-interarm oxygen abundance variations, in this study we analyse high spatial resolution MUSE data for a large sample of 45 spiral galaxies. To perform such a study, we compare the gas abundances of the spiral arms with those of the underlying disc (interarm region). In order to assess which factors may enhance or diminish the emergence of arm-interarm differences, we examine the dependence of the variations with three galaxy properties: the stellar mass, the presence of bars, and the flocculent/grand design appearance of spiral arms. Bars have been proposed as a key mechanism in the dynamical evolution of disc galaxies, for instance, by inducing gas flows \citep{lia1992, friedli1998}. We already mentioned that the existence of streaming motions along the spiral arms could cause the rise of azimuthal abundance variations as proved by numerical simulations, making the presence of a bar a relevant factor worth investigating. Based on our results, we discuss how the observed scenario could be explained according to the different theories on the nature of spiral structures.

The paper is organised as follows: Sec.~\ref{sec:sample} provides a description of the sample as well as the data used in this study. In Sec.~\ref{sec:analysis} we explain the methodology employed to detect the \hii regions, separate arm and interarm regions, and derive the corresponding oxygen abundances. The results of the analysis are presented in Sec.~\ref{sec:results}, including the study of the arm-interarm abundance differences, their dependence on different properties of the galaxies (Sec.~\ref{sec:properties}), and the comparison with a spaxel-wise analysis (Sec.~\ref{sec:spaxelwise}). The discussion and the main conclusions are given in Sec.~\ref{sec:dis}. Finally, a brief summary of the study is provided in Sec.~\ref{sec:summary}.

\section{Observations and galaxy sample}\label{sec:sample}
The analysed data belong to AMUSING++ (L\'opez-Cob\'a et al. submitted), a compilation of nearby galaxies observed with the MUSE instrument \citep{bacon2010, bacon2014}. This collection comprises 534 galaxies from different MUSE projects in combination with archival data covering the redshift interval $0.0002 < z < 0.1$. The core of the compilation comes from the AMUSING survey \citep[All-weather MUse Supernova Integral-field Nearby Galaxies; see][for more information on this survey]{galbany2016}, an ongoing project aimed at studying the environment of supernovae (SNe). In this work we make use of the 450 AMUSING galaxies observed up to Period 101 (September 2018) for which all data reduction, quality assessment, and sample characterisation has been performed (currently the project comprises about 600 galaxies). Observations for 362 of them constitute the long term observing campaign carried out as part of the AMUSING project, the remaining coming from the European Southern Observatory (ESO) archive with the only criterion of having hosted a SN. AMUSING++ complements this \mbox{AMUSING} sample by adding selected galaxies observed by MUSE not hosting SNe (see L\'opez-Cob\'a et al. submitted, for more details on the final AMUSING++ sample). Despite being the analysed dataset the result of a compilation, and therefore not comprising a homogeneously selected and well-defined sample, in L\'opez-Cob\'a et al. (submitted) the authors show that \mbox{AMUSING++} includes galaxies which properties resemble those of a diameter-selected sample. Not clear biases towards any particular morphological type, color, or magnitude are observed.

The MUSE instrument \citep{bacon2010, bacon2014} is mounted on the Unit 4 telescope (UT4) at the Very Large Telescope (VLT) in Cerro Paranal Observatory. In its Wide Field Mode, this integral-field spectrograph presents a FoV of approximately $1'\times1'$ and a pixel size of $0.2''$, which limits the spatial resolution of the data to the atmospheric seeing during the observations. Finally, MUSE covers a wavelength range from $\rm 4750$ to $\rm 9300$ \AA, with a spectral sampling of $\rm 1.25$ \AA\, and a spectral resolution $\lambda/\Delta\lambda \sim 1800-3600$ (from the blue edge to the red end of the spectrum).

A detailed explanation of the AMUSING data reduction is provided in \citet{kruhler2017}. Briefly, we make use of version 1.2.1 of the MUSE reduction pipeline \citep{weilbacher2014} and the Reflex environment \citep{freudling2013}. The sky subtraction is performed using the Zurich Atmosphere Purge package \citep[ZAP;][]{soto2016}, employing offset pointings to blank sky positions, or blank sky regions within the science frames. For the galaxies selected from the archive, the reduced cubes provided by ESO were used. The effects of Galactic extinction were also corrected based on the reddening estimates from \citet{schlegel1998}.

\vspace{0.5cm}
For this study, we select from the AMUSING mother sample a subset of 45 spiral galaxies. To reach that number we first remove galaxies with low image quality (visually presenting low signal in the source and/or with seeing values above $2 \arcsec$). In addition, observations that do not cover the centre of the galactic discs (hampering the derivation of required parameters such as the inclination or position angle of the galaxy) are also discarded from the sample. 

From a physical point of view, we select isolated spiral galaxies with morphological types between Sa and Sm (including barred galaxies) according to the {\it HyperLeda} extragalactic database\footnote{http://leda.univ-lyon1.fr} \citep{makarov2014}. For a proper separation of the arm and interarm regions, we restrict the sample to intermediate to low inclined galaxies ($i < 65 \degr$). Moreover, to perform a suitable characterisation of the abundance distribution of both areas, we discard those galaxies from the AMUSING sample that present a deprojected disc radius smaller than $\sim20 \arcsec$ and/or a radial coverage of less than one effective disc radius. In order to guarantee a feasible detection of the possible abundance variations, we only preserve galaxies for which a physical resolution higher than 1 kpc is achieved (determined from the seeing value of the observations). In addition, we also reject galaxies whose spiral arms can not be properly traced because of their poor definition (see Sec.~\ref{sec:arms} for details). These cases correspond to the most flocculent galaxies of the sample, whose fragmented and blurred arms are not easy to track, and galaxies with tightly wrapped ringlike arms, that are also very difficult to disentangle (see Sec.~\ref{sec:arms}). Finally, galaxies for which the \hii regions visually provide a poor spatial coverage of the disc (patchy and discontinuous) are also discarded. 

The required selection criteria ensure a proper characterisation of both the arm and interarm oxygen abundance distributions in our sample. However, due to these criteria the sample may not cover the full range of properties of the galaxy population in the Local Universe. In this sense, we must bear in mind that this study represents the first attempt to extend the analysis of arm-interarm abundance variations to a statistically significant sample. Having a complete sample was never the intention of the present work. Upcoming surveys massively sampling the Local Universe with higher spatial resolutions will enable our results to be placed within a broader framework.

For the final sample of 45 galaxies, we present in Table~\ref{table} a summary of the most important characteristics of the galaxies. Information on the redshift, physical resolution (limited by the seeing corresponding to the observations of each object), morphological type, arm classification, and stellar mass is provided in the table. 

\begin{table*}
\setlength{\tabcolsep}{10pt}
\renewcommand{\arraystretch}{1.1}
\caption{Fundamental properties of the galaxy sample.}
\label{table}
\begin{tabular}{l@{\hspace{1cm}}cccccccc}
\hline\hline\\
Name & Morph & z & Res. & $\log$ Mass & Arm & Bar & $\rm \Delta \,([O/H])_{HII}$ & $\rm \Delta \,([O/H])_{sp}$ \\[0.1cm]
  & type & & [pc] & [M$_{\odot}$] & class &  & [dex] & [dex] \\[0.2cm]
{\scriptsize(a)} & {\scriptsize(b)} & {\scriptsize(c)} & {\scriptsize(d)} & {\scriptsize(e)} & {\scriptsize(f)} & {\scriptsize(g)} & {\scriptsize(h)} & {\scriptsize(i)}\\[0.1cm]
\hline\\
2MASXJ01504127-1431032  & $ -1.0 $ & $ 0.034  $ & $ 868 $ & $ 10.8 $ &  F  &  U  & $ -0.006  \pm  0.008 $ & $ +0.002  \pm  0.007 $ \\
ESO~018--G018  & $ 4.2 $ & $ 0.021  $ & $ 528 $ & $ 11.4 $ &  GD  &  B  & $ -0.005  \pm  0.004 $ & $ +0.005  \pm  0.006 $ \\
ESO~184--G082  & $ 4.1 $ & $ 0.009  $ & $ 181 $ & $ 9.8 $ &  GD  &  B  & $ -0.036  \pm  0.015 $ & $ -0.025  \pm  0.005 $ \\
ESO~246--G021$^\star$  & $ 3.0 $ & $ 0.019  $ & $ 484 $ & $ 11.4 $ &  GD  &  B  & $ +0.012  \pm  0.009 $ & --- \\
ESO~467--G013  & $ 4.9 $ & $ 0.024  $ & $ 439 $ & $ 11.0 $ &  GD  &  U  & $ +0.001  \pm  0.007 $ & $ -0.003  \pm  0.005 $ \\
ESO~478--G006  & $ 4.2 $ & $ 0.018  $ & $ 462 $ & $ 11.3 $ &  GD  &  U  & $ +0.004  \pm  0.005 $ & $ +0.004  \pm  0.005 $ \\
ESO~498--G005$^\dagger$  & $ 4.3 $ & $ 0.008  $ & $ 123 $ & $ 10.5 $ &  GD  &  B  & $ +0.006  \pm  0.005 $ & --- \\
ESO~506--G004  & $ 2.5 $ & $ 0.013  $ & $ 863 $ & $ 11.0 $ &  ---  &  B  & $ +0.009  \pm  0.012 $ & $ +0.022  \pm  0.007 $ \\
ESO~570--G020  & $ 3.2 $ & $ 0.028  $ & $ 470 $ & $ 11.0 $ &  F  &  U  & $ +0.006  \pm  0.010 $ & $ +0.005  \pm  0.007 $ \\
IC~1320  & $ 2.9 $ & $ 0.017  $ & $ 449 $ & $ 10.7 $ &  F  &  B  & $ +0.022  \pm  0.007 $ & $ +0.006  \pm  0.006 $ \\
IC~2151  & $ 3.9 $ & $ 0.010  $ & $ 225 $ & $ 10.4 $ &  F  &  B  & $ +0.020  \pm  0.008 $ & $ +0.012  \pm  0.005 $ \\
MCG--01--57--021  & $ 4.0 $ & $ 0.010  $ & $ 263 $ & $ 10.7 $ &  GD  &  B  & $ -0.001  \pm  0.008 $ & $ +0.002  \pm  0.004 $ \\
MCG--04--38--04  & $ 5.5 $ & $ 0.032  $ & $ 682 $ & $ 11.7 $ &  GD  &  B  & $ +0.008  \pm  0.007 $ & $ +0.013  \pm  0.007 $ \\
NGC~0289$^\dagger$ & $ 4.0 $ & $ 0.005  $ & $ 68 $ & $ 10.7 $ &  GD  &  B  & $ +0.006  \pm  0.004 $ & --- \\
NGC~0835$^\star$  & $ 1.9 $ & $ 0.014  $ & $ 390 $ & $ 11.0 $ &  GD  &  B  & $ +0.004  \pm  0.007 $ & --- \\
NGC~0881$^\star$  & $ 5.0 $ & $ 0.018  $ & $ 517 $ & $ 11.2 $ &  GD  &  B  & $ +0.005  \pm  0.006 $ & --- \\
NGC~1080  & $ 4.7 $ & $ 0.026  $ & $ 744 $ & $ 11.3 $ &  GD  &  B  & $ +0.028  \pm  0.009 $ & $ +0.030  \pm  0.006 $ \\
NGC~1285  & $ 3.4 $ & $ 0.017  $ & $ 607 $ & $ 11.2 $ &  GD  &  B  & $ +0.015  \pm  0.011 $ & $ +0.001  \pm  0.005 $ \\
NGC~1309$^\dagger$  & $ 3.9 $ & $ 0.007  $ & $ 104 $ & $ 11.0 $ &  ---  &  U  & $ -0.009  \pm  0.004 $ & --- \\
NGC~1483$^\dagger$  & $ 4.0 $ & $ 0.004  $ & $ 101 $ & $ 10.4 $ &  F  &  B  & $ -0.052  \pm  0.011 $ & --- \\
NGC~1591  & $ 2.0 $ & $ 0.014  $ & $ 582 $ & $ 11.1 $ &  GD  &  B  & $ +0.001  \pm  0.005 $ & $ +0.003  \pm  0.006 $ \\
NGC~1762  & $ 5.1 $ & $ 0.016  $ & $ 335 $ & $ 11.4 $ &  GD  &  U  & $ +0.012  \pm  0.004 $ & $ +0.007  \pm  0.005 $ \\
NGC~2370  & $ 3.4 $ & $ 0.018  $ & $ 439 $ & $ 11.2 $ &  GD  &  B  & $ +0.007  \pm  0.005 $ & $ +0.016  \pm  0.006 $ \\
NGC~2466  & $ 5.0 $ & $ 0.018  $ & $ 490 $ & $ 11.2 $ &  F  &  U  & $ +0.008  \pm  0.005 $ & $ +0.004  \pm  0.004 $ \\
NGC~3120$^\dagger$  & $ 4.2 $ & $ 0.009  $ & $ 149 $ & $ 10.7 $ &  GD  &  B  & $ +0.014  \pm  0.006 $ & --- \\
NGC~3244  & $ 5.7 $ & $ 0.009  $ & $ 193 $ & $ 10.7 $ &  GD  &  U  & $ -0.006  \pm  0.005 $ & $ -0.006  \pm  0.003 $ \\
NGC~3318  & $ 3.7 $ & $ 0.009  $ & $ 275 $ & $ 10.9 $ &  GD  &  B  & $ -0.001  \pm  0.006 $ & $ +0.000  \pm  0.005 $ \\
NGC~3363  & $ 3.5 $ & $ 0.019  $ & $ 562 $ & $ 11.1 $ &  GD  &  U  & $ +0.018  \pm  0.007 $ & $ +0.007  \pm  0.006 $ \\
NGC~3464  & $ 4.9 $ & $ 0.012  $ & $ 338 $ & $ 11.1 $ &  GD  &  B  & $ +0.007  \pm  0.005 $ & $ +0.022  \pm  0.005 $ \\
NGC~3512$^\dagger$  & $ 5.1 $ & $ 0.005  $ & $ 138 $ & $ 10.4 $ &  GD  &  B  & $ +0.006  \pm  0.005 $ & --- \\
NGC~3783  & $ 1.4 $ & $ 0.010  $ & $ 200 $ & $ 11.1 $ &  GD  &  B  & $ +0.007  \pm  0.007 $ & $ +0.014  \pm  0.004 $ \\
NGC~3905  & $ 4.7 $ & $ 0.019  $ & $ 365 $ & $ 11.4 $ &  GD  &  B  & $ +0.041  \pm  0.007 $ & $ +0.042  \pm  0.006 $ \\
NGC~4981$^\dagger$  & $ 4.0 $ & $ 0.006  $ & $ 112 $ & $ 11.1 $ &  GD  &  B  & $ +0.001  \pm  0.005 $ & --- \\
NGC~5339  & $ 1.3 $ & $ 0.009  $ & $ 659 $ & $ 10.7 $ &  GD  &  B  & $ +0.002  \pm  0.005 $ & $ +0.020  \pm  0.007 $ \\
NGC~5584$^\dagger$  & $ 5.9 $ & $ 0.005  $ & $ 145 $ & --- &  GD  &  B  & $ -0.016  \pm  0.007 $ & --- \\
NGC~6708  & $ 2.6 $ & $ 0.009  $ & $ 203 $ & $ 10.9 $ &  GD  &  U  & $ +0.003  \pm  0.007 $ & $ +0.002  \pm  0.002 $ \\
NGC~6754  & $ 3.1 $ & $ 0.011  $ & $ 365 $ & $ 11.0 $ &  GD  &  B  & $ +0.031  \pm  0.006 $ & $ +0.033  \pm  0.004 $ \\
NGC~6806  & $ 5.0 $ & $ 0.019  $ & $ 691 $ & $ 11.3 $ &  GD  &  B  & $ -0.019  \pm  0.011 $ & $ -0.012  \pm  0.006 $ \\
NGC~7421$^\dagger$  & $ 3.7 $ & $ 0.006  $ & $ 118 $ & $ 10.8 $ &  GD  &  B  & $ +0.005  \pm  0.005 $ & --- \\
PGC~004701  & $ 3.9 $ & $ 0.018  $ & $ 450 $ & $ 10.7 $ &  F  &  U  & $ -0.003  \pm  0.008 $ & $ +0.003  \pm  0.007 $ \\
PGC~1015413  & --- & $ 0.014  $ & $ 419 $ & $ 10.0 $ &  F  &  U  & $ -0.010  \pm  0.014 $ & $ -0.038  \pm  0.008 $ \\
PGC~127886  & $ 10.0 $ & $ 0.024  $ & $ 528 $ & $ 10.5 $ &  F  &  --  & $ -0.032  \pm  0.009 $ & $ -0.038  \pm  0.005 $ \\
PGC~128348  & $ 5.0 $ & $ 0.015  $ & $ 317 $ & $ 10.4 $ &  F  &  U  & $ -0.033  \pm  0.009 $ & $ -0.024  \pm  0.005 $ \\
UGC~01395  & $ 3.1 $ & $ 0.017  $ & $ 500 $ & $ 11.0 $ &  GD  &  U  & $ +0.029  \pm  0.007 $ & $ +0.027  \pm  0.006 $ \\
UGC~11214  & $ 5.9 $ & $ 0.009  $ & $ 188 $ & $ 10.6 $ &  F  &  U  & $ -0.018  \pm  0.013 $ & $ -0.029  \pm  0.005 $ \\
\hline\hline
\end{tabular}
\justify
{\bfseries Notes.} Columns contain: (a) The galaxy name; (b) the morphological type according to the de Vaucouleurs system; (c) the redshift; (d) the physical resolution determined from the seeing value (in pc); (e) the logarithm of the integrated stellar mass in units of solar masses; (f) the arm class (F for flocculent, GD for grand design); (g) the presence of bar (B for barred, U for unbarred); (h) average magnitude of the arm-interarm abundance variations from the analysis of the individual \hii regions; (i) average magnitude of the arm-interarm abundance variations from the spaxel-wise analysis. \\
$^\star$ Spaxel-wise analysis not feasible due to a poor spatial coverage of the abundance distribution (patchy or discontinuous) or to a insufficient number of star-forming spaxels defining the interarm region.\\
$^\dagger$ Spaxel-wise analysis not feasible due to the high physical resolution of the original datacubes ($< 150$ pc).
\end{table*}

\section{Analysis}\label{sec:analysis}
\subsection{Measurement of emission lines with \scshape{Pipe3d}}\label{sec:pipe3d}
We make use of the fitting package FIT3D and {\scshape Pipe3d} \citep{sanchez2016a, sanchez2016b}, an IFS analysis pipeline developed to characterise the properties of both the stellar population and the ionised gas. 

Here we provide a brief outline of the procedure of fitting and subtracting the underlying stellar population and measuring the emission lines using FIT3D. The entire scheme together with other algorithms of {\scshape Pipe3d} are extensively described in \citet{sanchez2016a, sanchez2016b}. Briefly, FIT3D fits each spectrum by a linear combination of simple stellar population templates \citep[following][]{cidfernandes2013} after correcting for the appropriate systemic velocity and velocity dispersion, and also for the stellar dust extinction \citep{cardelli1989}. 

After the stellar component is subtracted, FIT3D measures the emission line fluxes by performing a multi-component fitting using a weighted moment analysis, as described in \citet{sanchez2016b}. The measured line fluxes covered by MUSE include H$\alpha$, H$\beta$, $[\oiii]\lambda4959$, $[\oiii]\lambda5007$, $[\nii]\lambda6548$, $[\nii]\lambda6584$, $[\sii]\lambda6717$, and $[\sii]\lambda6731$, among others. FIT3D provides the intensity, equivalent width (EW), systemic velocity, and velocity dispersion for each emission line of each spectrum. For further analysis, emission line intensities are corrected for dust attenuation making use of the extinction law from \citet{cardelli1989}, with $R_V = 3.1$, and the H$\alpha$/H$\beta$ Balmer decrement, considering the theoretical value for the unobscured H$\alpha$/H$\beta$ ratio of 2.86, which assumes a case B recombination ($T_e = 10\,000$ K, $n_e=100$ cm$^{-3}$, \citealt{osterbrock1989}). 

\subsection{Selection of \hii regions}\label{sec:hiiregions}
We detect clumpy regions of ionised gas in each galaxy, candidates for \hii regions, based on its H$\alpha$ intensity map using {\scshape \mbox{HIIexplorer}} \citep{sanchez2012b}. A first description of the algorithm is provided there, but here we use an updated version modified for high resolution data like MUSE, that was described in \citet{sanchezmenguiano2018}. Regions presenting $S/N < 3$ in H$\alpha$ emission are rejected. On average we detect $\sim200$ \hii region candidates per galaxy, thus a total of $8896$ regions in the full sample.

\begin{figure}
	\includegraphics[width=\columnwidth]{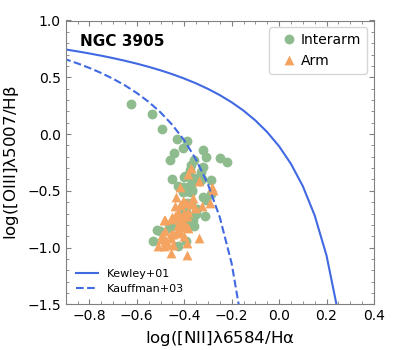}
    \caption{BPT diagram of the \hii regions in NGC~3905. The solid and dashed lines represent the \citet{kewley2001} and \citet{kauffmann2003} demarcation curves. Star-forming regions are considered to be below the \citet{kewley2001} curve.}
    \label{fig0}
\end{figure}

From these candidates, we select true star-forming \hii regions using the diagnostic BPT diagram proposed by \citet{baldwin1981}, based on the $[\nii]\lambda6584$/H$\alpha$ and $[\oiii]\lambda5007$/H$\beta$ line ratios. For this diagram we adopt the \citet{kewley2001} curve that separates regions ionised by OB stars (below the curve) from those associated with other sources of ionisation such as active galactic nuclei (AGNs) or shocks (above the curve). An additional criterion on H$\alpha$ EW (greater than 6 \AA) is also assumed to ensure the exclusion of low-ionisation sources \citep[such as weak AGNs or post-AGB stars,][]{cidfernandes2011}, and the presence of a significant percentage (at least $20\%$) of young stars contributing to the emission of the \hii regions \citep[given the strong correlation between both parameters, see][]{sanchez2014}. This procedure has proven to perform a good classification to describe real \hii regions, as shown by Espinosa et al. (in prep.).

Figure~\ref{fig0} shows the BPT diagram for one example galaxy in the sample, NGC~3905. We only represent the selected regions that are finally associated with star formation, distinguishing between those that belong to the spiral arms (orange) and the interarm area (green, see Sec.~\ref{sec:arms} for details in the separation of the arm and interarm areas). It is evident that both distributions spread differently across the BPT diagram, with the interarm regions presenting higher $\rm [\oiii]/H\beta$ line ratios than the arm regions (on average interarm values are 0.3 dex higher than arm values).

\subsection{Separation of arm and interarm regions}\label{sec:arms}
The procedure followed here to separate the arm and interarm regions was used in \citet{sanchezmenguiano2017} for CALIFA data. The outline of the spiral arms was based on previous and successful tracing of other morphological features such as dust lanes in galactic bars \citep[see][]{sanchezmenguiano2015}.

\begin{figure}
	\includegraphics[width=\columnwidth]{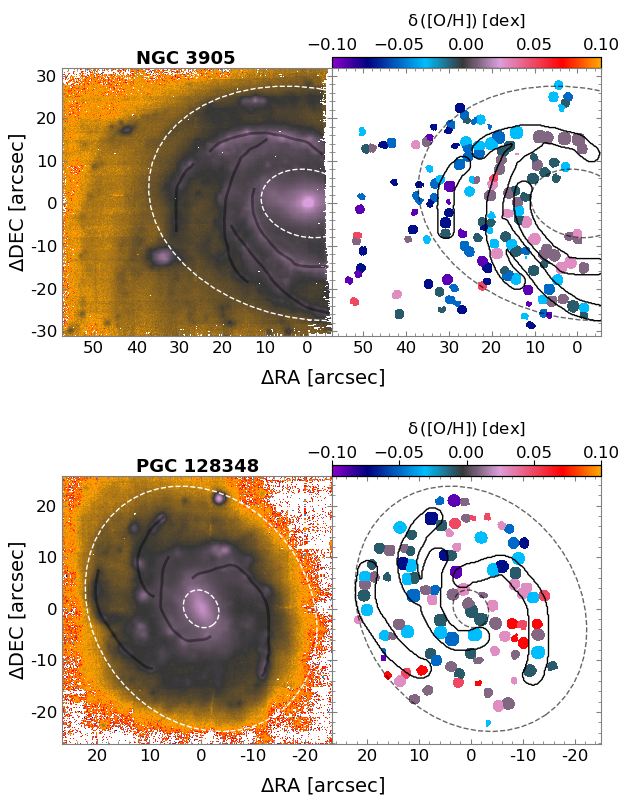}
    \caption{{\it Left panel:} Outline of the spiral arms on the $g$-band image for two example galaxies. {\it Right panel:} Colour map of the oxygen abundance residuals (derived by subtracting the radial profile) of the detected \hii regions. Black solid lines enclose the area defined as arm region. Dashed ellipses on both panels indicate the minimum and maximum radial limits (given by the minimum and maximum galactocentric distances of the arm region) to be considered for the interarm region.}
    \label{fig1}
\end{figure}

We first depict the spiral arms by visually tracing them on \mbox{$g$-band} images of the galaxies reconstructed from the datacubes. The marked points are then interpolated using a cubic spline (a numeric function that is piecewise-defined by a third-degree polynomial). When discontinuous arms are considered, we individually outline all fragmented parts that are easily distinguishable. The left panels of Fig.~\ref{fig1} show the spline fit of the detected spiral arms (dark grey solid lines) superimposed on the $g$-band images of two example galaxies, NGC~3905 (top) and PGC~128348 (bottom). Once the spiral structure is traced, we consider as arm \hii regions those separated from the closest point of the arms less than a certain distance that is visually chosen from the $g$-band image. The arm widths therefore range from 1 to 5\arcsec, that correspond to $0.2-1.6$ kpc \citep[in agreement with the values found in][]{honig2015, sanchezmenguiano2017}. The remaining \hii regions are associated with the interarm regions. The right panels of Fig.~\ref{fig1} show the division of arm and interarm \hii regions, being the ones belonging to the spiral arms those contained inside the line-delimited areas. The interarm \hii regions are those located outside the defined areas.

This method of tracing spiral arms restricts us to the most strong and prominent arms. Weaker spiral arms might be missed, being the \hii regions belonging to them wrongly associated to the interarm regions. This effect may dilute possible differences, and in this case the reported variations, if existing, should be considered as a lower limit to the real ones. Nevertheless, it is expected that the most prominent and easily tracked arms correspond to those having the strongest impact on the chemical distribution, producing the largest and most clearly observable abundance differences. 

\subsection{Derivation of oxygen abundances}\label{sec:ox} 
The wavelength range covered by MUSE (which does not include the $[\oii]\lambda3727$ emission line) prevents the use of several calibrators often adopted to measure oxygen abundances. This is the case for those based on the R23 or ONS indicators \citep[e.g.][]{zaritsky1994, kewley2002, pilyugin2010}. Among the remaining available calibrators, the ones based on the O3N2 index, defined as $\rm O3N2 = \log([\oiii]\lambda5007/H\beta \times H\alpha/[\nii]\lambda6584)$, stand out because of its monotonic dependence on the abundance and the close distance in wavelength between the lines of both ratios, which makes the index barely affected by dust attenuation. 

The O3N2 indicator was first introduced by \citet{alloin1979} and later modified by \citet{pettini2004}. In this study we derive the oxygen abundances of the \hii regions making use of this indicator in combination with a later calibration proposed by \citet{marino2013}, hereafter M13. By employing $T_e$-based abundances of $\sim 600$ \hii regions from the literature together with new measurements from the CALIFA survey, it constitutes one of the most accurate calibrations to date for the O3N2 index. The improvement of this calibration is especially significant in the high-metallicity regime, where previous calibrators based on this index lack high quality observations \citep[e.g.][]{pettini2004, perezmontero2009}.

When measuring oxygen abundances using strong-line indicators, one has to keep in mind the systematic discrepancies arising between the different proposed diagnostics and calibrations \citep[for an extended discussion, see][]{kewley2008, lopezsanchez2012}. For this reason, in order to ensure that the results are not contingent on the adopted method to derive the abundances we also make use of the \citet{dopita2013} calibration, hereafter D13, that is based on the MAPPINGS IV code developed by the authors. This calibration is based on a grid of photoionisation models, that cover a wide range of abundances and ionisation parameters typical of \hii regions in galaxies. D13 can be used through a Python module implemented by the authors, known as {\tt pyqz}, which is publicly available\footnote{\url{http://dx.doi.org/10.4225/13/516366F6F24ED}}. 

\begin{figure}
\centering
	\includegraphics[width=0.9\columnwidth]{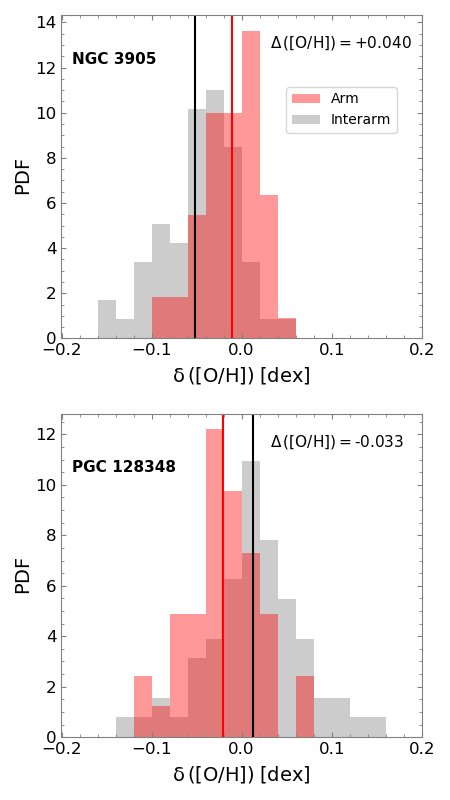}
    \caption{Probability distribution function (PDF) of the oxygen abundance residuals of the \hii regions for the two example galaxies shown in Fig.~\ref{fig1}. The red (grey) histogram correspond to the arm (interarm) distribution. The vertical solid lines represent the mean value for each distribution.}
    \label{fig2}
\end{figure}

For the sake of clarity, below we only show the results based on the use of the M13 calibrator. The results derived from the D13 diagnostic are provided in Appendix~\ref{app:pyqz}, showing that the main conclusions of the paper are independent of the adopted calibration, although slight differences arise. These differences will be discussed in Sec.~\ref{sec:dis}.

\begin{figure}
\centering
	\includegraphics[width=\columnwidth]{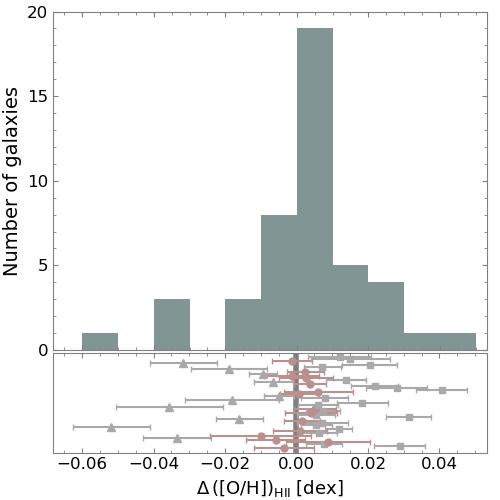}
    \caption{Distribution of average arm-interarm abundance variations for the galaxy sample derived from the analysis of the \hii regions. Individual values together with the errors are shown in the bottom panel. Red dots represent galaxies that are compatible with an absence of arm-interarm differences (vertical line centred at zero).}
    \label{fig3}
\end{figure}

\section{Results}\label{sec:results}
As introduced in Sec.\ref{sec:intro}, the principal behaviour observed in the oxygen abundance distribution of spiral galaxies is a negative radial gradient \citep[e.g.][]{smith1975, vilacostas1992, kennicutt2003, bresolin2012, sanchez2014, sanchezmenguiano2016, poetrodjojo2019}. Apart from this radial decline, other trends have been observed very often in the inner and outer parts of the discs, namely a decrease towards the centre of the galaxies \citep[e.g.][]{belley1992, rosalesortega2011, sanchez2014, sanchezmenguiano2016, zinchenko2016} and a flattening in the outskirts \citep[for a review, see][and references therein]{bresolin2017}. Following the procedure described in \citet{sanchezmenguiano2018}, we determine the characteristic radial profile of each galaxy in the sample, and then remove it in order to highlight the arm-interarm abundance variations. The so-called abundance residuals of the \hii regions ($\rm \delta \,([O/H])$) are therefore derived by subtracting the corresponding values for their galactocentric distances according to this radial trend to the observed ones. The abundances residuals range between $-0.25$ and $0.30$ dex, thus, clearly larger than the expected uncertainties for the considered calibrator\footnote{The derived abundances with M13 have a calibration error of $\pm 0.08$ dex, and the typical errors associated with the pure propagation of the errors in the measured emission lines are about 0.02 dex.}.

\begin{figure*}
\centering
	\includegraphics[width=2.1\columnwidth]{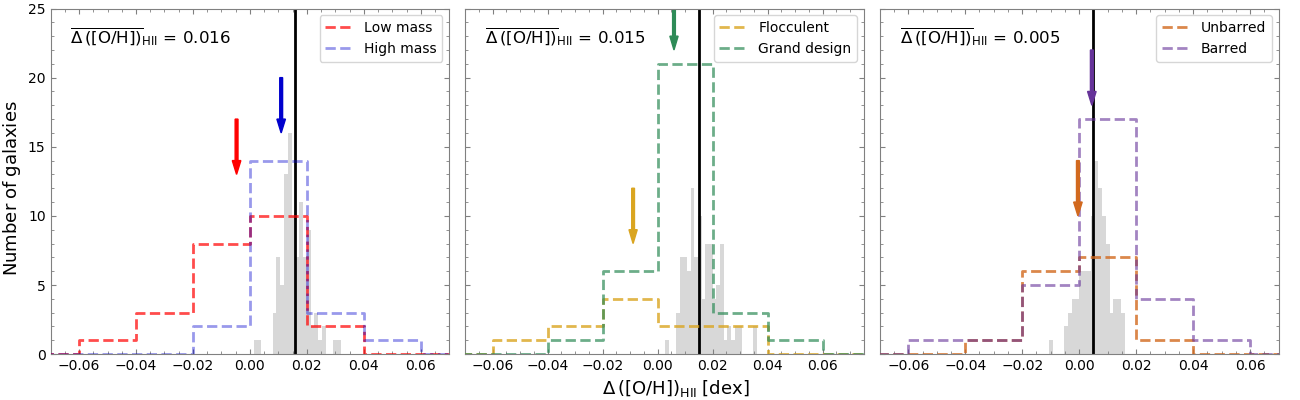}
    \caption{Arm-interarm abundance variations as a function of: galaxy mass ({\it left panel}), appearance of the spiral arms ({\it middle panel}), and presence of bars ({\it right panel}). Specific information for each panel is provided in the legend. Arrows represent the mean value of the distribution for each considered subset of galaxies. The result of the subtraction of both values is indicated in the top left corner of the panels and shown as black vertical lines. Grey shaded histograms represent a bootstrapped distribution of the difference in the average arm-interarm variations between the two split subsamples for each analysed parameter. A stellar mass cut of $10^{11} \, \rm M_\odot$ has been used.}
    \label{fig4}
\end{figure*}

The right panels of Fig.~\ref{fig1} display the map of oxygen abundance residuals of the \hii regions detected in two example galaxies. Distinguishing the regions that fall inside the line-delimited areas indicating the spiral arms from those located outside, we can derive separately the arm and interarm residual abundance distributions. Their probability distribution functions (PDF) are represented in Fig.~\ref{fig2} for the same example galaxies as red and grey histograms, respectively. For a fair comparison, we restrict the analysed interarm distribution to the \hii regions within the radial range covered by the spiral arms (defined by the dashed ellipses represented in Fig.~\ref{fig1}). Vertical lines of Fig.~\ref{fig2} represent the mean abundance residual of the arm ($\rm \overline{\delta \,([O/H])}_{\,arm}$) and interarm ($\rm \overline{\delta \,([O/H])}_{\,interarm}$) distributions. For the first galaxy, we can see how the \hii regions of the spiral arms are on average more metal rich than those belonging to the interarm area. The subtraction of the mean values of both distributions ($\rm \Delta \,([O/H])_{HII} = \overline{\delta \,([O/H])}_{\,arm} - \overline{\delta \,([O/H])}_{\,interarm}$) yields a difference of approximately $0.040$ dex. The second galaxy, in contrast, displays more metal-poor spiral arms compared with the rest of the disc ($\rm \Delta \,([O/H])_{HII} = -0.033$ dex).

The distribution of average arm-interarm abundance variations $\rm \Delta \,([O/H])_{HII}$ for the galaxy sample is shown in Fig.~\ref{fig3}. The individual values are listed in Table~\ref{table} and also represented with error bars in the bottom panel of the Figure. We find that 15 galaxies of the sample (\mbox{$\sim33\,\%$}, red dots) are compatible within errors with presenting no significant variations between the arm and interarm abundance distributions. In addition, 20 galaxies ($\sim45\,\%$, grey squares) display positive arm-interarm abundance variations, that is, spiral arms that are more metal rich than the interarm region. Finally, the remaining 10 galaxies ($\sim22\,\%$, grey triangles) show negative arm-interarm abundance variations, meaning that spiral arms are more metal poor than the interarm area in these systems. The magnitude of the reported differences is in all cases very small, with the values ranging from -0.06 to 0.05 dex. Nevertheless, we should not forget that these differences are averaged over the whole arm and interarm areas. Abundance variations of up to $0.2-0.3$ dex exit when individual \hii regions in both areas are considered.

\subsection{Dependence on galaxy properties}\label{sec:properties}

Figure~\ref{fig3} exposes the diversity of behaviours displayed by the galaxy sample, with some objects exhibiting a metal rich spiral structure with respect to the interarm region, and other objects showing the opposite trend. A homogeneous arm-interarm abundance distribution has also been revealed for some galaxies. The large number of systems comprising the sample allows to explore how the emergence of arm-interarm abundance variations may depend on different galaxy properties. Here we focus on three global parameters that may enable light to be shed on the origin of the spiral structure: the stellar mass, the appearance of the spiral arms (flocculent or grand design), and the presence of bars. The values of these parameters for the whole galaxy sample are collected in Table~\ref{table}.

The left panel of Fig.~\ref{fig4} shows the effect of the galaxy mass. Stellar masses were derived as in \citet{sanchezmenguiano2018} by applying the mass-luminosity ratio of \citet{bell2001} to the $g$ and $r$ apparent magnitudes recovered from the data (equations in \citealt{jester2005} were used to transform from SDSS magnitudes to the Johnson system). Galaxies have been split into two subsets: higher- (blue dashed histogram) and lower- (red) mass systems, using a cut of $10^{11} \, \rm M_\odot$ in order to have a comparable number of elements in both subsets. On average, more massive galaxies present larger positive $\rm \Delta \,([O/H])_{HII}$ values, that is, more metal rich spiral arms with respect to the interarm area. The difference in the mean $\rm \Delta \,([O/H])_{HII}$ values between higher-mass (blue arrow) and lower-mass (red arrow) systems is $0.016$ dex (vertical black line). Performing a bootstrapping scheme of the data (technique based on random sampling with replacement) 100 times (grey shaded histogram), we obtain that this difference is always systematically positive, concluding that the found dependence on the galaxy mass is not driven by extreme cases. Furthermore, we perform a two-sample Kolmogorov-Smirnov test (K-S test) to check if the differences found in the results between higher- and lower-mass galaxies are statistically significant. The resulting $P$-value is $0.1\%$, well below the significance level of 5 per cent, supporting the result that lower- and higher-mass systems have a different distribution of arm-interarm abundance variations. 

The second analysed parameter is the appearance of the spiral structure. In this way, galaxies are divided into flocculent and grand design systems according to the symmetry and continuity of the spiral pattern \citep{elmegreen1982, elmegreen1987}. Flocculent galaxies present small and patchy spiral arms while grand designs are characterised by the presence of long, symmetric and continuous arms. The classification was carried out based on a visual inspection of $g$-band images recovered from the data, resulting in 11 flocculent galaxies and 32 grand designs (two galaxies were ambiguous and therefore not considered). The middle panel of Fig.~\ref{fig4} shows the distribution of average arm-interarm abundance variations for flocculent (yellow dashed histogram) and grand design (green) systems. Although the covered range for both distributions is similar, on average the grand design galaxies display larger positive arm-interarm abundance variations (i.e. more metal rich spiral arms compared to the interarm area) than the flocculent ones. The difference in the mean $\rm \Delta \,([O/H])_{HII}$ values is $0.015$ dex (vertical black line). Again, the performed bootstrapping yields a distribution of values always above 0.0 dex, indicating that the arm-interarm abundance variations are systematically larger (positive) in grand design galaxies than in flocculent ones. This is confirmed by the $P$-value of $3\%$ obtained from the K-S test.

Finally, we investigate the role of the galaxy bar on the emergence and magnitude of the arm-interarm abundance variations, the outcome of which is shown in the right panel of Fig.~\ref{fig4}. Galaxies are separated in barred (purple dashed histogram) and unbarred (brown) systems according to the information provided in HyperLeda. One galaxy, for which this information is not available in the database, has been excluded. In this case, the distributions of average arm-interarm abundance variations for barred and unbarred galaxies are very similar, with a $\rm \overline{\Delta \,([O/H])}_{HII}$ value of $0.005$ dex. The performed bootstrapping, with a distribution of values compatible with zero, and the K-S test, with a $P$-value of $21\%$, confirm this result. 

In summary, our results seem to indicate that the chemical enrichment associated with the spiral arms is clearly correlated with the galaxy mass and the appearance of the spiral structure. More massive and grand design galaxies show spiral arms that are more metal rich than the interarm area compared with less massive and flocculent objects. This enrichment of the spiral arms seems to weaken until reversing, with some cases in which the spiral structure presents less metal rich gas than the interarm regions, preferentially observed in lower-mass and flocculent galaxies. The presence of a bar does not seem to play any role in this scenario, with no significant differences in the arm-interarm abundance distribution reported between barred and unbarred galaxies.

\subsection{Comparison with a spaxel-by-spaxel approach}\label{sec:spaxelwise}
A drawback of selecting \hii regions in order to characterise the gas abundance distribution is the limitation in the statistics and the coverage of the arm and/or the interarm areas. The advent of IFS techniques offers the opportunity to overcome these limitations by tracing the distribution of ionised gas using all the available information (spectra) across the entire galaxies extent. However, this approach suffers from other problems, such as possible dilution effects and an imprecise decontamination of the underlying diffuse nebular emission \citep{reynolds1984, oey2007, zhang2017}.

In order to reinforce the results derived analysing individual \hii regions, we carry out here an alternative analysis using a spaxel by spaxel approach. In this way, to derive the arm and interarm abundance distributions we follow the same procedure described in Sec.~\ref{sec:analysis} for the clumpy regions detected with {\scshape \mbox{HIIexplorer}}. However, in this case we base the analysis on all the spaxels from the datacubes that fulfil the criteria defined in Sec.~\ref{sec:hiiregions} to be associated with star formation.

The high spatial resolution of the data entails an obstacle to carry out a spaxel by spaxel analysis. Their physical resolution may allow us to resolve the ionised structure of the \hii regions in some galaxies, depending on the atmospheric seeing during the observations and the redshift of the object \citep[see further details in][]{kruhler2017}. This represents an issue since the empirical calibrators used to derive the oxygen abundances cannot be applied when resolving different areas of \hii regions \citep{lopezhernandez2013, terlevich2014}. To overcome this difficulty, we degrade the data to reach a physical resolution below the order of the typical size of a ``giant'' extragalactic \hii region \citep[a few hundred parsecs, e.g.][]{oey2003}. For this, we perform a simple spatial binning scheme (squared bins) in order to have a final spatial unit of \mbox{$\rm \sim 400 \,pc$}. However, when the original physical resolution of the datacubes is very high, we would have to add a lot of pixels to reach the final resolution of \mbox{$\rm \sim 400 \,pc$}, loosing significant spatial information when degrading the data. For this reason, we exclude from this analysis all galaxies that present an original physical resolution of the datacubes ($< 150$ pc). In addition, galaxies for which the analysed star-forming spaxels visually provide a poor spatial coverage of the disc (patchy and discontinuous) are also not considered (ESO~246-G021, NGC~0835, and NGC~0881). The final sample for the spaxel by spaxel analysis comprises 33 galaxies.

\begin{figure}
\centering
	\includegraphics[width=0.9\columnwidth]{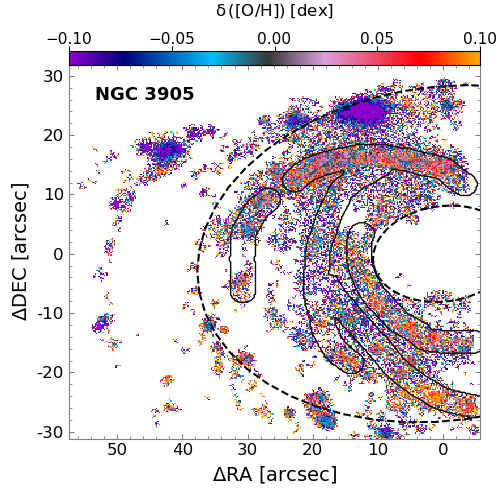}
    \caption{Colour map of the oxygen abundance residuals (derived by subtracting the radial profile) of the star-forming spaxels in NGC 3905. See caption of the right panels of Fig.~\ref{fig1} for more details.}
    \label{fig5}
\end{figure}

Figure~\ref{fig5} shows the spaxel-wise residual oxygen abundance distribution of NGC~3905, for which the same map based on the individual \hii regions is displayed in the top-right panel of Fig.~\ref{fig1}. In this map it can be clearly observed how the spiral arms present higher abundance residuals (orange-yellow colours) than the interarm region (blue-purple). Deriving the PDFs of the arm and interarm abundance distributions in a similar way as for the \hii regions, we find that NGC~3905 has an average arm-interarm abundance variation $\rm \Delta \,([O/H])_{sp}=0.03$ (compared to the 0.04 value obtained from the \hii region analysis). For the whole galaxy sample, we identify 14 galaxies with positive arm-interarm abundance variations ($43\%$), and 7 galaxies with negative variations ($21\%$). The remaining 12 galaxies ($36\%$) are compatible within errors with an absence of significant arm-interarm differences. The $\rm \Delta \,([O/H])_{sp}$ values for the galaxies are provided in Table~\ref{table}. We note that although the $\rm \Delta \,([O/H])$ values slightly change between the analysis of the \hii regions and the spaxel-by-spaxel approach, the qualitative behaviour regarding the presence of positive or negative arm-interarm variations (or the compatibility with an absence of them) is the same (taking into account the error intervals) for all galaxies except for a very small percentage of the sample (ESO 506–G004, IC~1320, NGC~1285, and NGC~5339).

\begin{figure*}
\centering
	\includegraphics[width=2.1\columnwidth]{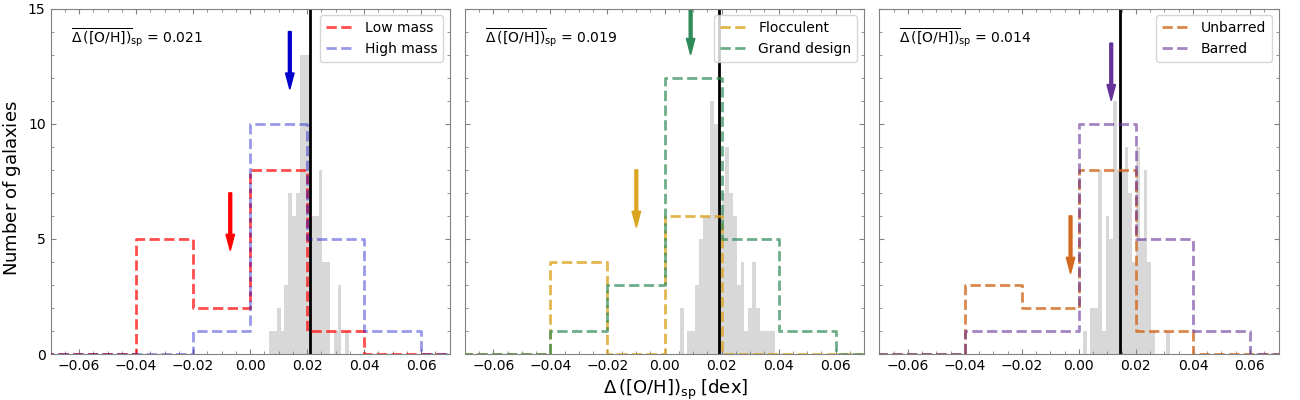}
    \caption{Spaxel-wise arm-interarm abundance variations as a function of: galaxy mass ({\it left panel}), appearance of the spiral arms ({\it middle panel}), and presence of bars ({\it right panel}). See caption of Fig.~\ref{fig4} for more details.}
    \label{fig6}
\end{figure*}

Finally, we investigate again the dependence of the arm-interarm abundance variations with the same three previously studied galaxy properties:  the stellar mass, the appearance of the spiral arms, and the presence of bars. The outcome of this test is shown in Fig.~\ref{fig6}. Analogously to when analysing the individual \hii regions, a correlation with the galaxy mass (left panel) appears, with more massive galaxies displaying more metal rich spiral arms compared to the interarm area. The difference in the mean $\rm \Delta \,([O/H])_{sp}$ values between higher-mass (blue arrow) and lower-mass (red arrow) systems is $0.021$ dex, a bit larger that the value obtained in the \hii region analysis. The significance of this difference is supported by the performed bootstrapping (positive distribution of values) and the K-S test ($P$-value of $0.5\%$). Similarly, when examining the effect of the appearance of the spiral structure, the spaxel by spaxel analysis reinforces the results from the \hii region one: grand design galaxies (green histogram) tend to present larger positive arm-interarm abundance variations than flocculent systems (yellow, $\rm \overline{\Delta \,([O/H])}_{sp} = 0.019$ dex). The bootstrapping confirms this trend, although in this case the $P$-value of the K-S test is above the significance level ($15\%$). It can not be discarded that the low number of flocculent systems (11) may affect the accuracy of the K-S test. Lastly, regarding the role of the galactic bar in shaping the arm-interarm abundance variations, the spaxel wise analysis yields larger differences between barred (purple) and unbarred (brown) galaxies ($\rm \overline{\Delta \,([O/H])}_{sp} = 0.014$ dex). The outcome of the performed bootstrapping, with a distribution of values always above zero, and the K-S test, with a $P$-value of $3\%$, confirm the statistical significance of this result. 

In summary, we have tried to improve the number statistics and the coverage of the arm and interarm areas by following a spaxel by spaxel approach for a smaller galaxy sample (33 of the 45 galaxies comprising the original sample). This analysis reinforces the results obtained with the classical approach of selecting individual \hii regions regarding the role of the galaxy mass and the appearance of the spiral structure in the emergence and magnitude of the arm-interarm abundance variations. In addition, despite the weak effect previously observed for the presence of a galactic bar, in this case the analysis shows significant arm-interarm abundance variations between barred and unbarred systems, with barred galaxies displaying spiral arms that are more metal rich than the interarm area compared with unbarred objects.

\section{Discussion}\label{sec:dis}

In the past, the chemical distribution of the ISM in spiral galaxies was considered to be highly homogeneous (\citealt{scalo2004}, see also \citealt{martin1996,cedres2002}). However, the advent of new IFS instruments combining high spatial resolution and large field-of-views has recently brought to light the presence of azimuthal variations on the gas oxygen abundance distribution in different galaxies \citep{sanchez2015a, sanchezmenguiano2016, vogt2017, ho2017, ho2018}. All these works point to a connection between the observed chemical inhomogeneities and the spiral structure of the galaxy. Although the above studies agree that the gas in the spiral arms is more metal-rich than the rest of the disc (interarm region), the magnitude of the reported variations diverges. The fact that the analysed data are collected using different instruments and that the employed methodologies vary, hampers a proper comparison of the arm-interarm abundance differences among the explored galaxies, and therefore, hinders the assessment of the physical scenario behind the observed trends. 

In this study we take advantage of the availability of high spatial resolution MUSE data for a large sample of galaxies within the AMUSING project \citep{galbany2016}. For 45 spiral galaxies (that remain from the initial sample after the selection criteria outlined in Sec.~\ref{sec:sample}), we compare the gas abundance distribution of the spiral arms with that of the interarm region in order to evaluate (with larger statistics) whether the spiral structure is the main driver of the abundance variations. We report the presence of more metal-rich gas in the spiral arms with respect to the interarm region for a subsample of galaxies ($\sim40-45\%$), confirming previous studies. Surprisingly, a small percentage of the sample ($\sim20\%$) shows the opposite trend, that is, spiral arms that are more metal poor than the interarm area. The remaining $\sim35-40\%$ of the galaxies do not display any clear enrichment pattern associated with the spiral arms, in agreement with \citet{kreckel2019}, that only find arm-interarm abundance variations in half of their sample. However, Fig.~\ref{fig3} shows that the PDF of the abundance variations clearly differs from the normal distribution centred at zero, which is confirmed by performing a Lilliefors test \citep[][this test evaluates the compatibility with a Gaussian distribution]{lilliefors1967} that yields a $P$-value of $3\%$. Despite this, the amplitude of the stated abundance variations are very small (up to 0.05 dex) when compared to previous studies, that report variations as large as 0.4 dex \citep{ho2017}. However, we note that these differences are averaged over the whole arm and interarm areas, whereas in most of previous studies the provided values represent the maxima of the detected variations. \citet{ho2018}, following a similar methodology to separate arm and interarm areas, report abundance differences in NGC~2997 of 0.02 dex when the same calibrator is used, totally compatible with the range of values obtained here. 

The obtained arm-interarm abundance variations rely on the defined width for the spiral arms. For this study, based on a thorough visual inspection of the $g$-band images of the galaxies, we measured arm widths ranging from $1-5\arcsec$ (see Sec.~\ref{sec:arms} for details). In order to check how the choice of this value may affect the results, we perform several tests using other values of the arm width in the analysis. When modifying the lower limit of $1\arcsec$ to $2\arcsec$, similar results are observed. However, as we increase the arm width the observed differences begin to disappear. This result is expected; by increasing the arm width, more interarm \hii regions are erroneously included in the spiral arms, and the arm-interarm abundance differences are diluted.

\begin{figure}
\centering
	\includegraphics[width=\columnwidth]{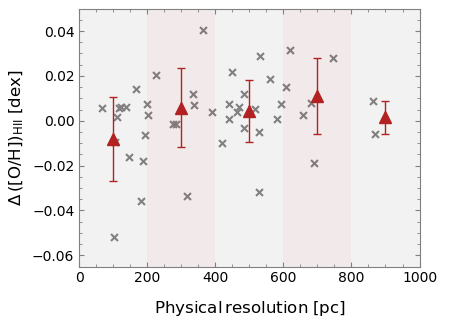}
    \caption{Average arm-interarm abundance variations for the galaxy sample as a function of the spatial resolution. Red triangles represent the mean values in bins of 200 pc, with the error bars indicating the standard deviation.}
    \label{fig7}
\end{figure}

Because of the nature of the AMUSING project, the observations used in this study present a wide variety of characteristics. In particular, the spatial resolution of the galaxies in the sample is very diverse, spanning from approximately 70 to 870 pc (see Table~\ref{table}). \citet{ho2018} addressed the issue of how the spatial resolution could affect the detectability of arm-interarm abundance differences. They concluded that the magnitudes of the variations decrease with worsening spatial resolution. However, this effect started to be noticeable with resolutions above 1 kpc, which is the upper limit selected for this study. Nevertheless, we represent in Fig.~\ref{fig7} the $\rm \Delta \,([O/H])$ values of the galaxies in the sample as a function of the spatial resolution. No clear trends are observed between both parameters, supported by a Pearson correlation coefficient of $r=0.27$ and a p-value for testing non-correlation of $8\%$ (i.e. above the significance level). This test suggests that the variety of detected arm-interarm abundance variations is not driven by differences in the spatial resolution of the data.

The large sample of galaxies analysed in this work enables the study of the dependency of the arm-interarm abundance variations with physical parameters of the galaxies. This analysis could help to identify which mechanisms may enhance or diminish the emergence of such variations. Here we focus in three galaxy properties: the stellar mass, the appearance of the spiral structure, and the presence of bars, which might provide key information about the nature of the spiral pattern. We observe that the first two factors seem to affect the chemical distribution of the galaxies, with more massive and grand design systems presenting larger positive arm-interarm abundance variations than their lower-mass and flocculent counterparts. This result is found when analysing the individual \hii regions and is reinforced by a second analysis following a spaxel-wise approach for a subsample of 33 galaxies. Regarding the role of the bar, a very weak enrichment of the spiral arms is observed in its presence when individual \hii regions are analysed, being significantly stronger in the spaxel-wise analysis. Further analysis based on a larger sample may help to evaluate whether the improvement in the number statistics and spatial coverage of the disc in the second approach is behind this detection, otherwise elusive (\hii region analysis).

In order to assess if the three studied galaxy properties act independently in shaping the gas abundance distribution we check whether there is any correlation between these parameters or not. Regarding the influence of bars, we find no clear correlation between the presence of a bar and any of the other analysed attributes (i.e. the stellar mass of the galaxy or the flocculent/grand design appearance of the spiral structure). This suggests that galactic bars themselves do seem to influence the chemical distribution of the host galaxies, by favouring the enrichment of the gas associated with the spiral arms. As regards the two remaining factors, the larger positive differences found in grand design and higher-mass galaxies might be related, since both properties seem to be connected. \citet{elmegreen2011}, analysing a sample of 46 galaxies from the S$^4$G project, observed a tendency in which early types (generally more massive) have grand design spirals, while the late types have flocculent ones. In our case, this behaviour is not so clear, but we find that higher-mass galaxies tend to present grand design spiral arms (at lower masses we find both flocculent and grand design arms), and a flocculent pattern is preferentially found in lower-mass galaxies (whereas grand design arms are equally found in lower- and higher-mass galaxies). We would like to note that although the sample has been divided in lower- and higher- mass galaxies in order to investigate the effect of the galaxy mass on the results, the used cut is $10^{11}$ M$_\odot$ (to have a comparable number of elements in both subsets). In general the sample comprises quite massive systems with values of log M$_\odot$ above $9.8$ (see Table~\ref{table}), which could influence the lack of correlation with the presence of a bar or the weak one observed with the appearance of the spiral pattern. A larger sample with a better coverage of the parameter space, including the low mass range, would be needed in order to corroborate these tendencies and disentangle the effect of both the galaxy mass and the appearance of the spiral structure in the chemical distribution of the galaxies.

The dependence of the found arm-interarm abundance variations with global galaxy properties is not new. \citet{sanchezmenguiano2017} analysed the radial abundance gradient in a sample of 63 galaxies from the CALIFA survey differentiating between arm and interarm regions. The authors reported very subtle differences in the observed gradients when segregating the sample in different subgroups, with the ionised gas in the inter-arm regions exhibiting a shallower gradient with a lower zero-point value and a larger dispersion in the oxygen abundances compared to that of the spiral arms. These differences displayed by the two areas were only observed for barred and flocculent galaxies. This is in agreement with our current results that larger (positive) arm-interarm abundance variations exist in barred galaxies with respect to unbarred systems, but does not agree with the larger (positive) variations observed in this work for grand design galaxies compared to flocculent systems. However, the different approaches followed in both studies (in \citealt{sanchezmenguiano2017} they study radial abundance gradients whereas here we directly compare the distribution of abundance values) does not allow us to further investigate the origin of the discrepancies. More recently, \citet{sanchezmenguiano2019} studied local spatial variations of the gas oxygen abundance distribution (as a proxy of the gas metallicity Z$_g$) using a large sample of $\sim 700$ galaxies from the MaNGA survey \citep{bundy2015}, based on data of much lower spatial resolution than in this study. Significant chemical inhomogeneities up to 0.2 dex were revealed across the discs, that were also found to correlate with local variations of star formation rate. The authors showed that the slope of the SFR-Z$_g$ correlation depended on the average gas-phase metallicity of the galaxy and its stellar mass. More metal-poor (and low mass) galaxies displayed the lowest slopes (i.e. the strongest SFR-Z$_g$ anticorrelations), reversing the relation for more metal-rich (and high mass) systems (at $\rm \sim 10^{10.5}-10^{11} M_\odot$). Since the spiral arms of galaxies are usually associated with regions of enhanced star formation rate \citep[e.g.][]{molla2019, spitoni2019}, the positive correlation found between SFR and Z$_g$ for massive systems is somehow analogous to the positive arm-interarm abundance variations found in this study. In the same way, the anticorrelation displayed by low mass galaxies is comparable to the negative arm-interarm abundance variations also revealed by our results. Although the number of systems displaying positive correlations (and equivalently, positive arm-interarm abundance variations) was lower in \citet{sanchezmenguiano2019} than in this study, it can be explained by the dominance of low mass galaxies in their sample. The SFR-Z$_g$ correlation was suggested to be motivated by external gas accretion. Indeed, it could be the case that in low mass galaxies the spiral arms are channels of pristine gas (i.e. metal-poor) coming from the intergalactic medium, whereas in high mass systems the spiral arms are dominated by localised metal recycling by preexisting gas. 

The arm and interarm abundance distributions compared in this study are determined making use of the empirical calibration proposed by \citet{marino2013} for the O3N2 indicator. In addition, the analysis is also reproduced using an alternative calibrator described in \citet{dopita2013} based on photoionisation models (see Appendix~\ref{app:pyqz}). Both analyses lead to equivalent results despite their origin being quite distinct, which reflects the robustness of the results independently of the adopted method to derive the oxygen abundances. The dependence of the arm-interarm abundance variations with the explored galaxy properties maintain. The only significant difference that arises from the use of these two calibrators is the number of galaxies displaying spiral arms that are more metal poor than the interarm area. Whereas the percentage of galaxies presenting negative $\rm \Delta \,([O/H])$ values is as high as $20\%$ for the M13 calibrator, this number decreases to barely $5\%$ in the case of D13. This could indicates that the two calibrators are affected in a different way by their relations with the ionisation parameter \citep[e.g.][]{morisset2016}, that is, the shape of the ionising spectrum. Thus, the differences or lack of them could be tracing changes not entirely related with the oxygen abundance distribution. Further analysis would be necessary in order to assess the impact of the ionisation parameter on the results. Thus, although we can not confirm the existence of galaxies with metal-poor spiral arms (compared to the interarm region), the opposite behaviour is strongly supported. From this study it is clear that the abundance distribution in spiral galaxies is not completely homogeneous and that the spiral structure plays a significant role in the local chemical enrichment of these systems, which somehow is also affected by the galaxy mass, the presence of a bar, and the appearance of the spiral pattern.

A first plausible explanation for the distinct oxygen abundance distributions shown by the ionised gas in the spiral arms and the underlying disc could suggest the presence of \hii regions of a different nature. \citet{kennicutt1989b}, and later \citet{ho1997}, reported the existence of certain \hii regions that distinguish themselves from the `classical' ones due to a stronger low-ionisation forbidden emission. Although first associated with the centres of galaxies, they were later found at any galactocentric distance \citep{sanchez2014}. The involvement of other sources of ionisation such as shocks were proposed to explain the origin of these \hii regions, although other stellar processes such as ageing were also suggested to produce the same effects \citep{sanchez2014}. Independently of their origin, this type of \hii regions is less easily detected in spiral arms due to its low number compared with that of classical \hii regions and the crowding of the later in this area. For this reason, they could be responsible for the arm-interarm abundance variations found in this study. However, these \hii regions have been found to present enhanced $[\nii]\lambda6584$/H$\alpha$ ratios and lower $[\oiii]\lambda5007$/H$\beta$ values, and to occupy the region in the BPT diagram between the \citet{kewley2001} and \citet{kauffmann2003} demarcation lines \citep{ho1997}. As we can see in Fig.~\ref{fig0}, this is not the case for the \hii regions linked to the interarm area in NGC~3905, which exhibit higher $[\oiii]\lambda5007$/H$\beta$ ratios and cover the same range of $[\nii]\lambda6584$/H$\alpha$ values than the \hii regions belonging to the spiral arms. They could still be responsible for the negative arm-interarm abundance variations found in a small subsample of galaxies, but they are insufficient to explain the whole picture.

Besides this possibility, the most plausible explanation for the detected arm-interarm abundance variations is that the physical processes behind these variations are connected to the dynamics of the gas and the spiral pattern. Recent 2D chemical evolution models that incorporate the role of the spiral pattern have shown how spiral density waves associated with this structure can be responsible of producing azimuthal abundance variations \citep{molla2019, spitoni2019}. The azimuthal trend would be the result of an enhanced star formation rate linked to an increase in the gas surface density due to the passage of the density wave. This effect would be strengthen by a raise in the probability of cloud-cloud collisions because of the produced shocks when the gas enters the arm \citep[e.g.][]{kobayashi2007}. However, the predicted abundance variations decrease significantly with time, especially for the SWS and SWR models presented in \citet{molla2019} and that of \citet{spitoni2019}, the three including the rotation of both the disc and spiral pattern. In these cases the mixing of the gas is higher due to rotation, diluting very quickly the azimuthal differences. As a consequence, the fact that we are able to observe clear arm-interarm abundance variations implies that the spiral arms had to be formed recently (less than 1-2 Gyr before the observations). Another alternative is that the production of the spiral pattern is a recurrent process along the evolution of the disc and we are observing the effects of the last created density wave. In this framework, the different reported amplitudes of the arm-interarm abundance variations could be the result of different time intervals from the formation of the spiral arms. 

Except for the above mentioned chemical evolution models, there are few theoretical works centred on the effect of spiral arms on the metallicity distribution of galaxies. Moreover, the existing studies are focused on the stellar metallicity rather than the behaviour of the gas. \citet{dimatteo2013}, analysing N-body simulations, have shown the existence of significant azimuthal variations in the metallicity distribution of old stars as a consequence of the effect of radial migration along spiral arms induced by a bar over a pre-existing radial metallicity gradient. Also based on N-body simulations, \citet{martinezmedina2016} found evidences of radial migration induced by both a galactic bar and the spiral structure producing variations in the stellar metallicity distribution with respect to the radial gradient \citep[see also][]{martinezmedina2017}. \citet{grand2016}, using high-resolution cosmological simulations, exposed an azimuthal variation in the stellar metallicity driven by streaming motions of star particles along the leading and trailing sides of the spiral arms. These radial flows produce an overdensity of metal-rich stars on the trailing edge of the spiral and metal-poor stars on the leading edge. Regarding the effect of the spiral arms on the gas metallicity distribution, \citet{sanchezmenguiano2016b} compared their observations on NGC~6754 with an N-body simulation showing how streaming motions of gas along spiral arms could also produce the same trends reported by \citet{grand2016} on the gas oxygen abundance distribution. The analysed simulations described spiral arm morphological features that were transient and rotated at a similar speed as the gas at every radius \citep[see][]{sellwood2011}, proposing a particular scenario for spiral structure formation of NGC~6754. On the contrary, \citet{ho2017} stated that the action of radial flows alone could not explain the high abundance variations observed in NGC~1365, and proposed another scenario within the density wave paradigm based on a simple chemical evolution model. According to this scenario, the arm-interarm abundance variations observed in the galaxy were caused by self-enrichment undergone by gas when crossing the interarm region followed by a mixing-induced dilution by the passage of spiral density waves. 

Based on the conclusions reached by previous works, we find that the complementarity of the analysis of the chemical distribution presented here with a future analysis on the gas velocity distribution for the same galaxy sample could be key in order to investigate the existence of radial flows in these galaxies and their connection with the observed arm-interarm abundance distributions. Furthermore, the analysis on the velocity pattern would enable the derivation of the corotation radius location, in order to confirm theoretical predictions finding stronger abundance variations close by these regions \citep{spitoni2019}. Both aims will be addressed in a forthcoming paper. In addition, more theoretical works focused on the response of the gas abundance distribution to the presence of the spiral structure would be crucial in order to investigate the effect of spiral arms of different nature \citep[e.g. steady arms caused by density waves, transient arms formed through local instabilities, or bar driven spirals; see review by][]{dobbs2014}. Of particular interest would be the predictions on the behaviour of the arm-interarm abundance variations as a function of the galaxy properties analysed in this study in order to distinguish between different spiral arm models. In this regard, simulations covering galaxies with different values of these properties (in particular, low mass and flocculent galaxies) could enable the finding of systems with more metal-poor spiral arms, allowing us, among other things, to confirm or discard the existence of negative arm-interarm abundance variations.

\section{Summary}\label{sec:summary}
In this work we compare the arm and interarm oxygen abundance distributions using high resolution IFS MUSE data for a sample of 45 spiral galaxies in order to assess the role of the spiral structure on the chemical enrichment of galaxies. We follow two complementary methodologies: one based on the analysis of the individual \hii regions, and another one using the information from all the spaxels associated with star formation. Whereas the first approach is more successful dealing with dilution effects and the decontamination of the underlying diffuse ionsed gas, the latter allows us to improve the statistics and the spatial coverage of the explored regions. Regarding the derivation of the oxygen abundances, we make use of two independent methods: an empirical calibrator based on the O3N2 indicator \citep[M13;][]{marino2013}, and a theoretical calibrator based on photoionisation models \citep[D13;][]{dopita2013}.

Comparing the abundance distribution of the arm and interarm regions of the galaxies, we find more metal-rich \hii regions in the spiral arms with respect to the interarm area for a large subsample of galaxies ($45\%$ using the M13 calibration, $65\%$ for D13). This is in agreement with previous studies on individual galaxies \citep{sanchezmenguiano2016, vogt2017, ho2017, ho2018}. Particularly relevant is the compatibility found between the range of values obtained here (up to $0.08$ dex) and the one derived by \citet{ho2018} for NGC~2997 following a similar methodology. In addition, surprisingly, we observe the opposite trend in a small percentage of the sample, that is, more metal-poor \hii regions in the spiral arms compared to the interarm region. This finding is highly dependent on the used calibrator, with $\sim 20\%$ galaxies exhibiting this behaviour for M13, and barely $5\%$ for D13, and therefore, further analysis is required to confirm it.

We investigate the dependence of the arm-interarm abundance variations with three galaxy properties: the stellar mass, the flocculent/grand design appearance of the spiral structure, and the presence of galactic bars. Following the two described approaches, we observe that the arm-interarm abundance differences are larger (positive) in more massive and grand-design galaxies than in low-mass and flocculent systems. In addition, the spaxel-wise analysis yields also significant differences when analysing the effect of bars, with barred galaxies presenting larger (positive) abundance variations. These trends are found irrespective of the used calibrator. 

The results suggest that the detected abundance variations are connected to the dynamics of the gas and the spiral structure. Further analysis on the gas velocity distribution could be useful in order to link these variations with the presence of radial flows, if existing. In addition, predictions from theoretical works on the response of the gas abundance distribution to the action of the spiral structure, as well as how the arm-interarm variations would depend on the analysed galaxy properties, are important to interpret the observations and distinguish between different spiral arm models.

\section*{Acknowledgements}
We would like to thank the anonymous referee for comments which helped to improve the paper. We are grateful to Merce Romero and Mercedes Moll\'a for their valuable comments. LSM acknowledges financial support from the Spanish Ministerio de Ciencia, Innovaci\'on y Universidades (MCIU) via grant AYA2016-79724-C4-2-P. SFS is grateful for the support of the CONACYT grants CB-285080 and FC-2016-01-1916, and funding from the PAPIIT-DGAPA-IA101217 (UNAM) project. TRL acknowledges financial support through the grants (AEI/FEDER, UE) AYA2017-89076-P, AYA2016-77237-C3-1-P and AYA2015-63810-P, as well as by the MCIU, through the State Budget and by the Consejer\'\i a de Econom\'\i a, Industria, Comercio y Conocimiento of the Canary Islands Autonomous Community, through the Regional Budget. TRL is supported by a MCIU Juan de la Cierva - Formaci\'on grant (FJCI-2016-30342). LG was funded by the European Union's Horizon 2020 research and innovation programme under the Marie Sk\l{}odowska-Curie grant agreement No. 839090. This study was based on observations made with ESO Telescopes at the Paranal Observatory (programmes 60.A-9317(A), 60.A-9392(A), 095.B-0532(A), 095.D-0172(A), 095.D-0091(B), 096.B-0309(A), 096.B-0951(A), 096.D-0263(A), 096.D-0296(A), 097.B-0640(A), 097.D-0408(A), 098.D-0115(A), 099.D-0022(A), 100.D-0341(A), and 0101.D-0748(B)). This research makes use of python (\url{http://www.python.org}), of Matplotlib \citep[][]{hunter2007}, a suite of open-source python modules that provides a framework for creating scientific plots, and Astropy, a community-developed core Python package for Astronomy \citep[][]{astropy2013}.




\bibliographystyle{mnras}
\bibliography{bibliography} 



\appendix

\section{Results with \tt{pyqz}}\label{app:pyqz}
The analysis described throughout the paper has been carried out making use of the O3N2 calibration from \citet[][M13]{marino2013} as the main indicator to derive the arm and interarm oxygen abundance distributions. In this appendix we assess how the results of the study may change by using other method to measure the oxygen abundance. Here we focus on the calibration proposed by \citet{dopita2013} based on photoionisation models (hereafter D13, see Sec.~\ref{sec:ox} for a brief description of this additional calibrator).

\begin{figure}
\centering
	\includegraphics[width=\columnwidth]{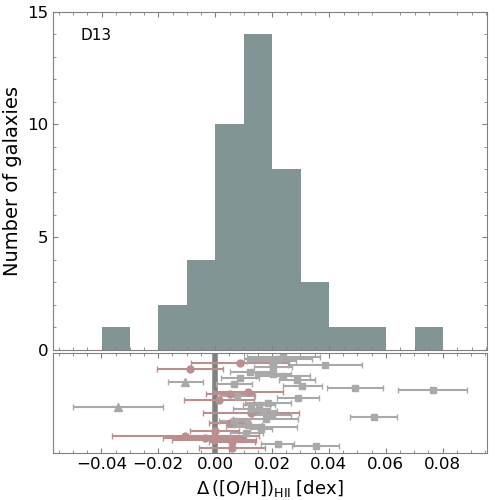}
    \caption{Distribution of average arm-interarm abundance variations for the galaxy sample measured using the D13 calibration  \citep{dopita2013}. Individual values together with the errors are shown in the bottom panel. Red dots represent galaxies that are compatible with an absence of arm-interarm differences (vertical line centred at zero).}
    \label{figapp1}
\end{figure}

Figure~\ref{figapp1} shows the distribution of average arm-interarm abundance variations $\rm \Delta \,([O/H])_{HII}$ using the D13 calibration for all the galaxies in the sample (based on the analysis of the individual \hii regions). Again, we can see that the distribution is shifted from the zero value, reflecting a non-homogeneous abundance distribution. However, unlike the analysis based on the M13 calibration, in this case most galaxies display positive arm-interarm abundance variations (30, $67\%$). The number of galaxies presenting negative $\rm \Delta \,([O/H])_{HII}$ values is very low, just 2 ($4\%$). The remaining 13 objects ($29\%$) are compatible within errors with an absence of significant differences between the arm and the interarm abundance distributions. The magnitude of the reported differences are slightly larger than for the M13 abundance indicator, ranging between -0.04 and 0.08 dex. Although the percentages slightly change when following the spaxel-wise approach, the trends are exactly the same: the majority of the galaxies exhibit positive arm-interarm abundance variations ($85\%$), and very few of them are associated with more metal poor spiral arms ($6\%$).

\begin{figure*}
\centering
	\includegraphics[width=2.1\columnwidth]{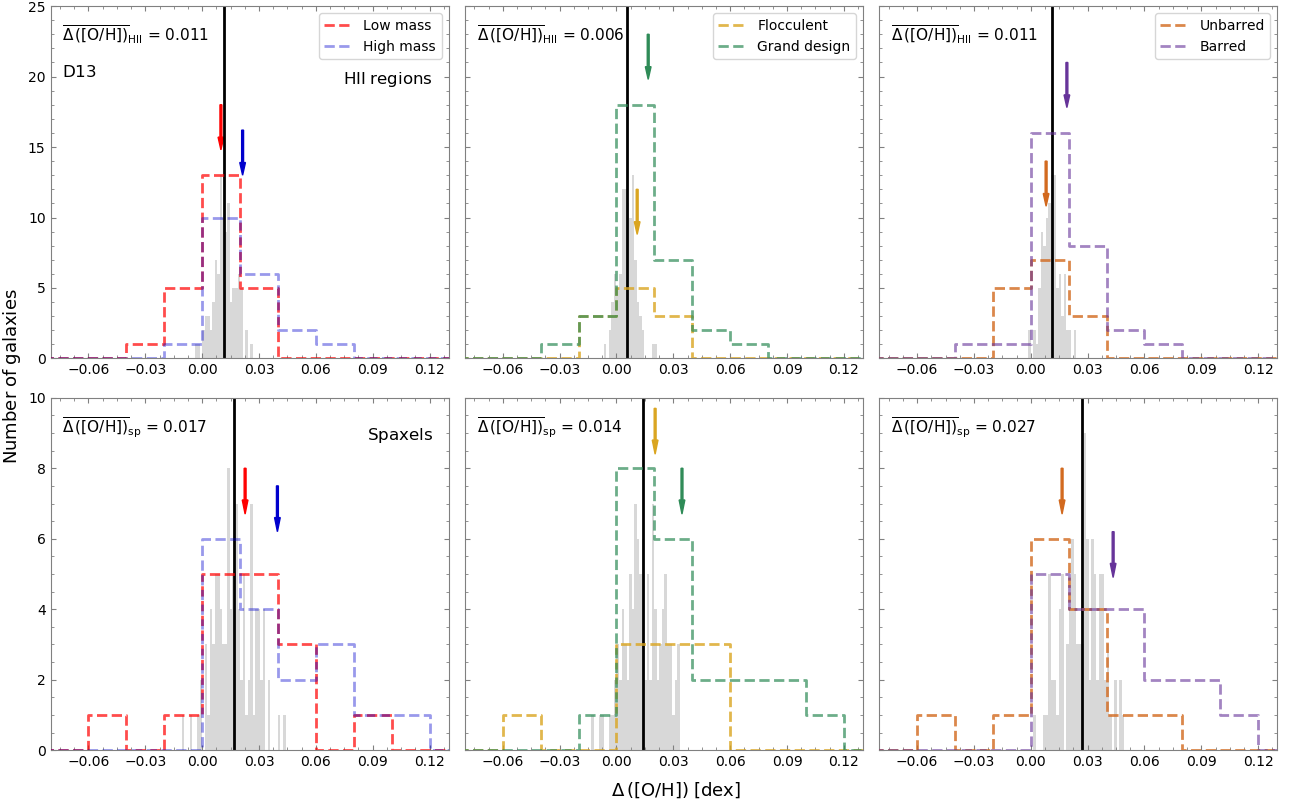}
    \caption{Arm-interarm abundance variations derived using the D13 calibrator from an \hii region-based ({\it top panels}) and spaxel-wise ({\it bottom panels}) analysis as a function of: galaxy mass ({\it left panels}), appearance of the spiral arms ({\it middle panels}), and presence of bars ({\it right panels}). See caption of Fig.~\ref{fig4} for more details.}
    \label{figapp2}
\end{figure*}

The arm-interarm abundance variations measured with D13 follow the same trends than those from M13 when investigating their dependence with the galaxy properties. Figure~\ref{figapp2} shows how more massive, grand design, and barred galaxies present larger positive arm-interarm abundance variations than low-mass, flocculent, and unbarred systems. As in the case of M13, the spaxel-by-spaxel analysis yields larger differences than the one based on the selection of individual \hii regions.


\bsp	
\label{lastpage}
\end{document}